# Entangling Independent Photons by Time Measurement

Matthäus Halder, Alexios Beveratos, Nicolas Gisin, Valerio Scarani, Christoph Simon & Hugo Zbinden

*Group of Applied Physics, University of Geneva, 20, rue de l'Ecole-de-Médecine, 1211 Geneva 4, Switzerland*

**A quantum system composed of two or more subsystems can be in an entangled state, i.e. a state in which the properties of the global system are well defined but the properties of each subsystem are not. Entanglement is at the heart of quantum physics, both for its conceptual foundations and for applications in information processing and quantum communication. Remarkably, entanglement can be "swapped": if one prepares two independent entangled pairs A1-A2 and B1-B2, a joint measurement on A1 and B1 (called a "Bell-State Measurement", BSM) has the effect of projecting A2 and B2 onto an entangled state, although these two particles have never interacted or shared any common past[1,2]. Experiments using twin photons produced by spontaneous parametric down-conversion (SPDC) have already demonstrated entanglement swapping[3-6], but here we present its first realization using continuous wave (CW) sources, as originally proposed[2]. The challenge was to achieve sufficiently sharp synchronization of the photons in the BSM. Using narrow-band filters, the coherence time of the photons that undergo the BSM is significantly increased, exceeding the temporal resolution of the detectors. Hence pulsed sources can be replaced by CW sources, which do not require any synchronization[6,7], allowing for the first time the use of completely autonomous sources. Our experiment exploits recent progress in the time precision of photon detectors, in the efficiency of photon pair production by SPDC with waveguides in nonlinear crystals[8], and in the stability of narrow-band filters. This approach is independent of the form of entanglement; we employed time-bin**

**entangled photons[9] at telecom wavelengths. In addition to entangling photons from autonomous sources, a fundamental quantum phenomenon, our setup is robust against thermal or mechanical fluctuations in optical fibres thanks to cm-long coherence lengths. The present experiment is thus an important step towards real-world quantum networks with truly independent and distant nodes.**

The BSM is the essential element in an entanglement-swapping experiment. Linear optics allows the realization of only a partial BSM[10] by coupling the two incoming modes on a beam-splitter (BS) and observing a suitable detection pattern in the outgoing modes. Such a measurement is successful in at most 50% of the cases. Still, a successful partial BSM entangles two photons that were, up to then, independent. The physics behind this realization is the bosonic character of photons, it is therefore crucial that the two incoming photons are indistinguishable: they must be identical in their spectral, spatial, polarization and temporal modes at the BS: Spectral overlap is achieved by the use of similar filters, spatial overlap by the use of single-mode optical fibres and polarization is matched by a polarization controller. In addition, the temporal resolution must be unambiguous: detection at a time $t \pm \Delta t_d$, with $\Delta t_d$ the temporal resolution of the detector, must single out a unique time mode. In previous experiments, synchronised pulsed sources created both the photons at the same time and path lengths had to be matched to obtain the required temporal overlap. The pulse length, i.e. the coherence length of the photons, was $\tau_c \ll \Delta t_d$ (typically $\tau_c <$1ps), but two subsequent pulses were separated by more than $\Delta t_d$[11]. The drawback of such a realization is that the two sources cannot be totally autonomous, because of the indispensable synchronization. Here, by using stable narrow-band filters and detectors with low jitter, we reach the regime where $\tau_c > \Delta t_d$[12]. In this case, the detectors always single out a unique time mode. As a benefit, we can give up the pulsed character of the sources and the synchronization between them, realizing for the first time the entanglement swapping scheme as originally proposed in Ref.2.

The experimental scheme is sketched in Fig.1. Each of the two non-linear crystals emits pairs of energy-time entangled photons[13] produced by SPDC of a photon originating from a CW laser. A pair can be created at any time $t$, and all these processes are coherent within the km-long coherence length of the laser: $|\psi\rangle_A \propto \sum_t |t,t\rangle_A$ describes a pair of signal and idler photons emitted by source A. Thus, the state produced by two independent sources can conveniently be represented as

$$|\Psi\rangle_{prep} = |\psi\rangle_A |\psi\rangle_B \propto \sum_t \left[ |t,t\rangle_A |t,t\rangle_B + \sum_{\tau>0} \left( |t,t\rangle_A |t+\tau,t+\tau\rangle_B + |t+\tau,t+\tau\rangle_A |t,t\rangle_B \right) \right].$$

The first term in the above sum describes 4 photons all arriving at the same time $t$ at a BS. Since for this case two identical photons bunch in the same mode, due to their bosonic nature, this term leads to a Hong-Ou-Mandel (HOM) dip[14]. The second term describes two photon pairs arriving with a time difference $\tau>0$. The two photons A1 and B1 are sent through a 50/50 BS. This fibre-coupler and the two detectors behind it realize a partial BSM[10]: in particular, when one of the detectors fires at time $t$ and the other one at time $t+\tau$, this corresponds to a measurement of the Bell-state $\Psi^-$ for A1 and B1[9]. In consequence the remaining two photons A2 and B2 are projected in the state $|\psi\rangle_{A2B2} \equiv |\Psi^-\rangle_{A2B2} \propto |t\rangle_{A2} |t+\tau\rangle_{B2} - |t+\tau\rangle_{A2} |t\rangle_{B2}$, which is a singlet state for time-bin entanglement. Hence entanglement has been swapped. This process can be seen as teleportation[15-19] of entanglement. It can be tested by sending the photons in unbalanced interferometers such that the path difference between the two arms corresponds to $\tau$. Interference between temporally distinguishable events (at $t$ and $t+\tau$, respectively) is obtained by erasing the time information via unbalanced interferometers[9,12,20] as shown in Fig.1. Note that the value of $\tau$ varies from one successful entanglement swapping event to another. As in our experiment the path differences of the analysing interferometers are fixed, we test the entanglement of the swapped pairs produced with one fixed $\tau$.

We now describe our experiment in more detail. Above we have assumed that the detection times $t$ and $t+\tau$ of the BSM are sharply defined. In physical terms, this requirement means that the detection times have to be determined with sub-coherence-time precision: this is the key ingredient that makes it possible to achieve synchronisation of photons A1 and B1 by detection, thus to use CW sources. Since single-photon detectors have a certain intrinsic minimal jitter, the coherence length of the photons has to be increased to exceed this value by narrow filtering.

Consider the case where each of the two sources emits one entangled pair of photons, and where A1 and B1 take different exits of the BS. The photon that takes output port 1 is detected by a NbN superconducting single-photon detector (SSPD)[21] with a time resolution $\Delta t_d = 74$ps. The photon in output port 2 is detected by an InGaAs single photon avalanche diode (APD, $\Delta t_d = 105$ps) triggered by the detection in the SSPD. The time resolution of these detectors is several times smaller than that of commercial telecommunication photon detection modules. To enable synchronization of the photons at the BS by post-selection, the coherence length of the photons has to exceed $\Delta t_d$. This is achieved by using filters of 10pm bandwidth, corresponding to a coherence time $\tau_c$ of 350ps. We are able to tolerate the losses due to filtering because we use cm-long wave guides in PPLN crystals with a high down-conversion efficiency of $5*10^{-7}$ per pump photon and per nm of the created spectrum. For 2mW of laser power, an emission flux $q$ of $2*10^{-2}$ pairs per coherence time is obtained. This $q$ is independent of the filtered bandwidth: in fact, narrower filtering decreases the number of photons per second but increases their coherence time by the same factor, hence keeping $q$ constant.

Any two-detector click in the BSM prepares the two remaining photons in a time-bin entangled state. In our experiment the creation rate for such entangled photon pairs is $\approx 10^4$ per second, with time delays $\tau$ ranging up to 10ns. This is two orders of magnitude larger than in previous experiments at shorter and similar wavelengths[3-6]. As the probability of both the pairs originating from different sources equals the probability of creating them in the same source, the first cases have to be post selected by considering only 4-fold-events. Furthermore only one fixed τ is tested. The resulting rate is smaller by two orders of magnitude compared to the creation rate. To verify their entanglement, the two photons are sent through unbalanced interferometers (*a* and *b*) in Michelson configuration. The path length differences of the interferometers must be identical only within the coherence length of the analyzed photons (7cm), but stable in phase (*α* and *β*): this is achieved by active stabilization[22]. On each side, both output ports of the interferometer are connected to InGaAs APD, triggered by the detection of both the photons in the BSM.

Four-fold coincidences, between one click in each BSM detector and one behind each interferometer, are registered by a multistop time to digital converter (TDC) and the arrival times (t, t+τ) are stored in a table. For τ = 0, we observe a decrease in this coincidence count rate (see Fig.2). The visibility of this HOM dip of 77% indicates the degree of indistinguishability of the two photons A1 and B1 and could be further improved by increasing $\tau_c/\Delta t_d$. The width of the dip corresponds to the convolution of $\tau_c$ for the two photons with the jitter of the detectors. Note that photons which are detected after the BS at measurable different times, but within $\tau_c$, do still partially bunch, which confirms that the relevant time precision is set by the coherence time of the photons.

To test for successful entanglement swapping, the relative phase *α-β* between the interferometers is changed by keeping *α* fixed and scanning *β*. As usual for the analysis of time-bin entanglement[9], interference is observable in the case where, at the output of the interferometers, both photons are detected at the same time.

We measured the four possible 4-fold coincidence count rates $R_{ij}(\alpha,\beta)$ (clicks in two outer detectors conditioned on a successful BSM) with $i,j \in \{+,-\}$ the different detectors behind interferometer *a* and *b*, respectively. Thus the two-photon spin-correlation coefficient $E(\alpha,\beta) = \frac{R_{++}(\alpha,\beta) - R_{+-}(\alpha,\beta) - R_{-+}(\alpha,\beta) + R_{--}(\alpha,\beta)}{R_{++}(\alpha,\beta) + R_{+-}(\alpha,\beta) + R_{-+}(\alpha,\beta) + R_{--}(\alpha,\beta)}$ is obtained as a function of the phase settings $\alpha$ and $\beta$ and plotted in Fig.3 for $\alpha$ fixed. A fit of the form $E(\alpha,\beta) = V\cos(\alpha-\beta)$ to our experimental data gives a visibility $V$=0.63. If one assumes that the two photons are in a Werner state (which corresponds to white noise), one can show that $V > \frac{1}{3}$ is sufficient to demonstrate entanglement[5,23]. Our experimental visibility clearly exceeds this bound[24]. The plain squares show that the 3-fold coincidence count rate between a successful BSM and only one of the outside detectors is independent of the phase setting, as expected for a $\Psi^-$-state. *V* is limited by imperfections in the matching of wavelengths, polarisations and temporal synchronisation. In our setup, the latter is the main source of errors. The integration time of this measurement was 1 hour for each of the 13 phase settings and the experiment was run 8 times, hence took 104 hours, which demonstrates the stability of our setup. Such long integration times are necessary because of low count rates (5 four-fold coincidences per hour), which are mainly due to poor coupling efficiencies of the photons into optical fibres, losses in optical components like filters and interferometers, as well as the limited detectors efficiencies. All these factors decrease the probability of detecting all four photons of a two-pair event.

Exploiting all the produced entangled pairs with different delays τ is possible in principle using rapidly adjustable delays in the interferometers or quantum memories. This would be an important step towards the realization of recent proposals for long-distance quantum communication[25]. Time-bin entanglement is particularly stable and well suited for fibre optic communications[26], and the coherence length of 7cm allows tolerating significant fiber length fluctuations as expected in field experiments. If additionally, count rates are further improved, long distance applications like quantum relays[27,28] become realistic.

In conclusion, we realized an entanglement swapping experiment with completely autonomous CW sources. This is possible thanks to the low jitter of new NbN superconducting and InGaAs avalanche single-photon detectors and to the long coherence length of the created photon pairs after narrow-band filtering. The setup does not require any synchronization between the sources and is highly stable against length fluctuations of the quantum channels.

**Methods**

Schematic description of the setup. Both sources consist of an external cavity diode laser in CW mode at 780.027nm (Toptica DL100), stabilized against a Rubidium transition ($^{85}$Rb F = 3), pumping a nonlinear periodically poled Lithium Niobate waveguide[8] (PPLN, HC photonics Corp) at a power of 2mW. The process of SPDC creates $4*10^{11}$ pairs of photons per second with a spectral width of 80nm FWHM centered at 1560nm. The photons are emitted collinearly and coupled into a single-mode fiber with 25% efficiency and the remaining laser light is blocked with a silicon high-pass filter (Si). Signal and idler photons are separated and filtered down to a bandwidth of 10pm by custom-made tunable phase-shifted Bragg gratings (AOS GmbH). These filters have a rejection of >40dB, 3dB insertion losses, and can be tuned independently over a range of 400pm. Once a signal photon has been filtered to $\omega_s$, the corresponding idler photon has a well-defined frequency $\omega_i$, due to stabilized pump wavelength and energy conservation in the process of SPDC ($\omega_s + \omega_i = \omega_{laser}$). After filtering, the effective conversion efficiency for creating a photon pair within these 10pm is $5*10^{-9}$ per pump photon. In principle, the available pump power permits us to produce narrow band entangled photon pairs at rates up to $3*10^8$ pairs per second, which translates to an emission flux of more than 0.1 photons per coherence time. In this experiment, we limited the laser to 2mW, in order to reduce the probability of multiple pair creation which would decrease the interference visibility[29].

After the beam splitter (BS), the first photon is detected by a NbN superconducting single-photon detector (SSPD, Scontel) operated in free running mode[21], with a total detection efficiency of 4.5%, 300 dark counts/sec and a timing resolution of 74ps, including the time jitter of both the detector and the amplification and discrimination electronics. The second photon is detected by an InGaAs single-photon avalanche diode operated in Geiger mode and actively triggered by the detection in the SSPD. With home-made electronics this detector has a time jitter of 105ps. The observed HOM-dip with a visibility of 77% was obtained with two SSPD detectors, which were used because of their smaller time jitter. For the entanglement swapping, we used an APD, because of its higher efficiency, in order to shorten the integration time. This means that the visibility of the interference fringe in Fig.3 could further be increased by the use of two SSPDs, but with the drawback of longer measurement times.

Photons A2 and B2 are also detected by InGaAs APDs (ID200, idQuantique). All the APDs have quantum efficiencies of 30% and dark count probabilities of $10^{-4}$ per ns. The interferometers are actively stabilized against a laser locked on an atomic transition, have a path length difference of 1.2ns and insertion losses of 4dB each.

Acknowledgements: We thank C. Barreiro, J.-D. Gautier, G. Gol'tsman, C. Jorel, S Tanzilli and J. van Houwelingen for technical support, and H. de Riedmatten, S. Iblisdir and R. Thew for helpful discussions. Financial support by the EU projects QAP and SINPHONIA and by the Swiss NCCR Quantum Photonics is acknowledged.

Author Information: Reprints and permissions information is available at npg.nature.com/reprintsandpermissions. The authors declare that they have no competing financial interests Correspondence and requests for materials should be addressed to M.H. (matthaeus.halder@physics.unige.ch).

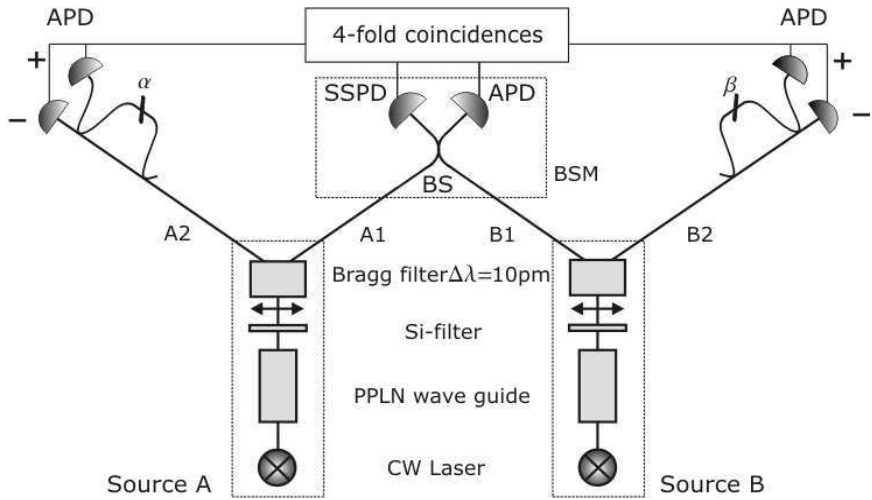

Figure 1: Experimental Setup. Two pairs of entangled photons (A1-A2 and B1-B2) are produced, one by each source (A and B), and all the photons are narrowly filtered (10pm). One photon of each pair is sent into a 50/50 beam splitter (BS) and both undergo a partial Bell-State measurement (BSM). By detection in different output ports of BS and with a certain time delay τ the two photons A1 and B1 are projected on the $\Psi^-$-state for time bin qubits, projecting the two remaining photons on the $\Psi^-$-state as well. The entanglement is swapped onto the photons A2 and B2 and can be tested by passing them through interferometers with phases α and β, and detecting them by single photon avalanche detectors (APD) in both outputs (+,-) of each interferometer.

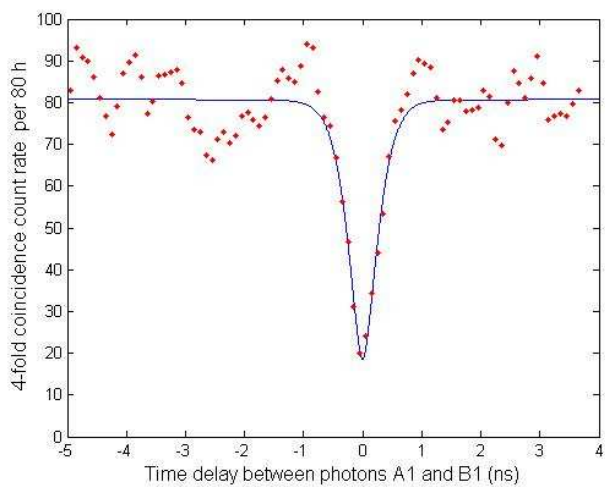

Figure 2: 4-fold coincidence count rate as a function of the temporal delay τ. It can be seen, that the detection probability decreases if the two photons A1 and B1 arrive simultaneously (τ=0) at the beam splitter due to photon bunching, leading to a Hong-Ou-Mandel dip with 77% visibility.

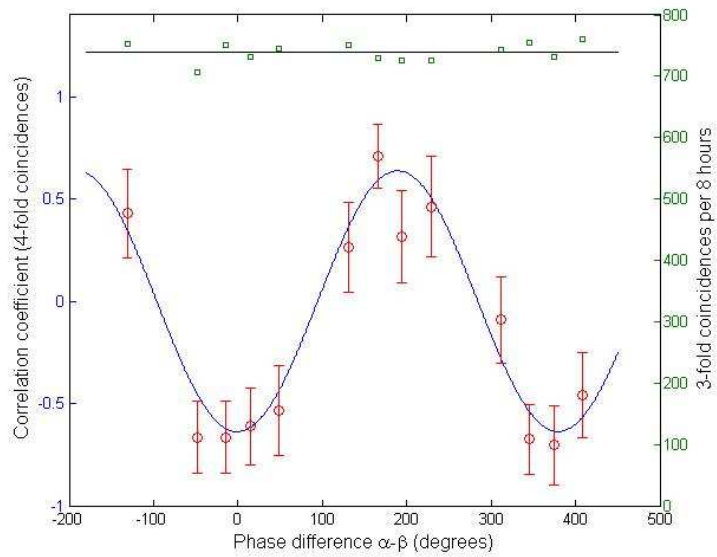

Figure3

Figure 3: The correlation coefficient E(α,β) between photons A2 and B2, conditioned on a BSM of photons A1 and B1, as a function of the relative phase α–β of the interferometers (open points). A fit of the form $E(\alpha,\beta) = V\cos(\alpha - \beta)$ gives a visibility V=0.63. This proves successful entanglement swapping (see text). The coincidence count rate of only one detector conditioned on a successful BSM (3-fold coincidence) is independent of the phase setting as expected for a $\Psi^-$-state (squares).